\documentclass[letter,aps,reprint,superscriptaddress]{revtex4-1}
\usepackage{amsmath}
\usepackage{braket}
\usepackage{graphicx}
\usepackage{textgreek}
\usepackage{xcolor}
\usepackage{mathtools}
\usepackage{ifthen}
\usepackage[normalem]{ulem}

\usepackage{listofitems}
\newcommand\substr[3]{%
  \setsepchar{#2}%
  \readlist\parsedinput{#1}%
  \foreachitem\x\in\parsedinput{%
    \ifnum\xcnt=1\else#3\fi\x%
  }%
}
\newcommand{\unit}[2]{#1\,\ifmmode{\mathrm{\substr{#2}{\mu}{\textmu}}}\else{\substr{#2}{\mu}{\textmu}}\fi}
\renewcommand{\H}{\hat{\mathcal{H}}}

\newcommand{\ETH}{Institute for Quantum Electronics, Eidgen\"ossische Technische Hochschule Z\"urich, Otto-Stern-Weg 1, 8093 Zurich, Switzerland}

\begin{document}

\title{P-band induced self-organization and dynamics with repulsively driven ultracold atoms in an optical cavity}
\date{\today}

\author{P.~Zupancic} \affiliation{\ETH}
\author{D.~Dreon} \affiliation{\ETH}
\author{X.~Li} \affiliation{\ETH}
\author{A.~Baumg\"artner}\affiliation{\ETH}
\author{A.~Morales} \affiliation{\ETH}
\author{W.~Zheng} \affiliation{T.C.M.~Group, Cavendish Laboratory, J.J.~Thomson Avenue, Cambridge CB3 0HE, United Kingdom}
\author{N.~R.~Cooper} \affiliation{T.C.M.~Group, Cavendish Laboratory, J.J.~Thomson Avenue, Cambridge CB3 0HE, United Kingdom}
\author{T.~Esslinger} \affiliation{\ETH}

\author{T.~Donner} \email{donner@phys.ethz.ch}   \affiliation{\ETH}

\begin{abstract}
We investigate a Bose-Einstein condensate strongly coupled to an optical cavity via a repulsive optical lattice. We detect a stable self-ordered phase in this regime, and show that the atoms order through an antisymmetric coupling to the P-band of the lattice, limiting the extent of the phase and changing the geometry of the emergent density modulation. Furthermore, we find a non-equilibrium phase with repeated intense bursts of the intra-cavity photon number, indicating non-trivial driven-dissipative dynamics.
\end{abstract}

\maketitle

Strong coupling between ultracold matter and quantized light fields can be achieved by placing a quantum gas in a high-finesse optical cavity. Their interplay generates non-linear atom-field dynamics that forms the basis for exploring collective many-body phenomena at the interface between quantum optics and condensed matter physics \cite{Ritsch2013,Gopalakrishnan2009,Mueller2012,Piazza2014,Keeling2014,Chitra2015,Schuetz2016,Himbert2019,Kirton2018,Li2013,Soriente2018}. A central phenomenon in this approach is the self-organization of a Bose-Einstein condensate (BEC) in a cavity mode \cite{Baumann2010}, when the atoms are illuminated from the side by a red-detuned, attractive standing wave pump lattice. Above a critical lattice depth, the particles are dragged into the intensity maxima of an emerging interference potential, thereby maximizing the scattering from the pump lattice to the cavity mode. The self-organization process is a second order phase transition at which the atoms reduce their potential energy in the modified potential landscape by a larger amount than the kinetic energy of the additional crystalline structure costs. This self-consistent ordering of atoms and light has become an experimental model system for driven-dissipative quantum phases \cite{Baumann2010,Klinder2015,Klinder2015a,Landig2016,Leonard2017,Kollar2017,Hruby2018,Landini2018,Kroeze2018,Vaidya2018}.

Theoretical studies have made predictions for possible self-ordered phases in a parameter regime where a quantum gas is coupled to a cavity via a blue-detuned optical lattice, which repels atoms from the intensity maxima \cite{Keeling2010,Ni2011,Bhaseen2012,Mivehvar2017}. Naively one could expect that self-organization is prohibited, since the buildup of any additional repulsive potential seems to cost energy. However, since the light field scattered by the atoms into the cavity is out of phase from the pump lattice field, destructive interference occurs at the position of the atoms. This carves out parts of the repulsive pump lattice potential, lowers the potential energy, and makes self-organization also possible for a blue-detuned pump (see Fig.~1). In such a configuration, limit cycles and chaos were recently predicted \cite{Keeling2010,Bhaseen2012,Diver2014,Piazza2015} and connected to dynamic phenomena such as time crystals \cite{Kessler2019}. Experiments, however, have so far been limited to the case of red-detuned, attractive pump fields. In this letter we report on self-organization with a repulsive optical lattice and identify the parameter regime for a stable phase, both experimentally and theoretically. We further discover experimentally a dynamic region, where the intra-cavity light field shows repeated pulses. 

\begin{figure}
\centering
\includegraphics[width=\columnwidth]{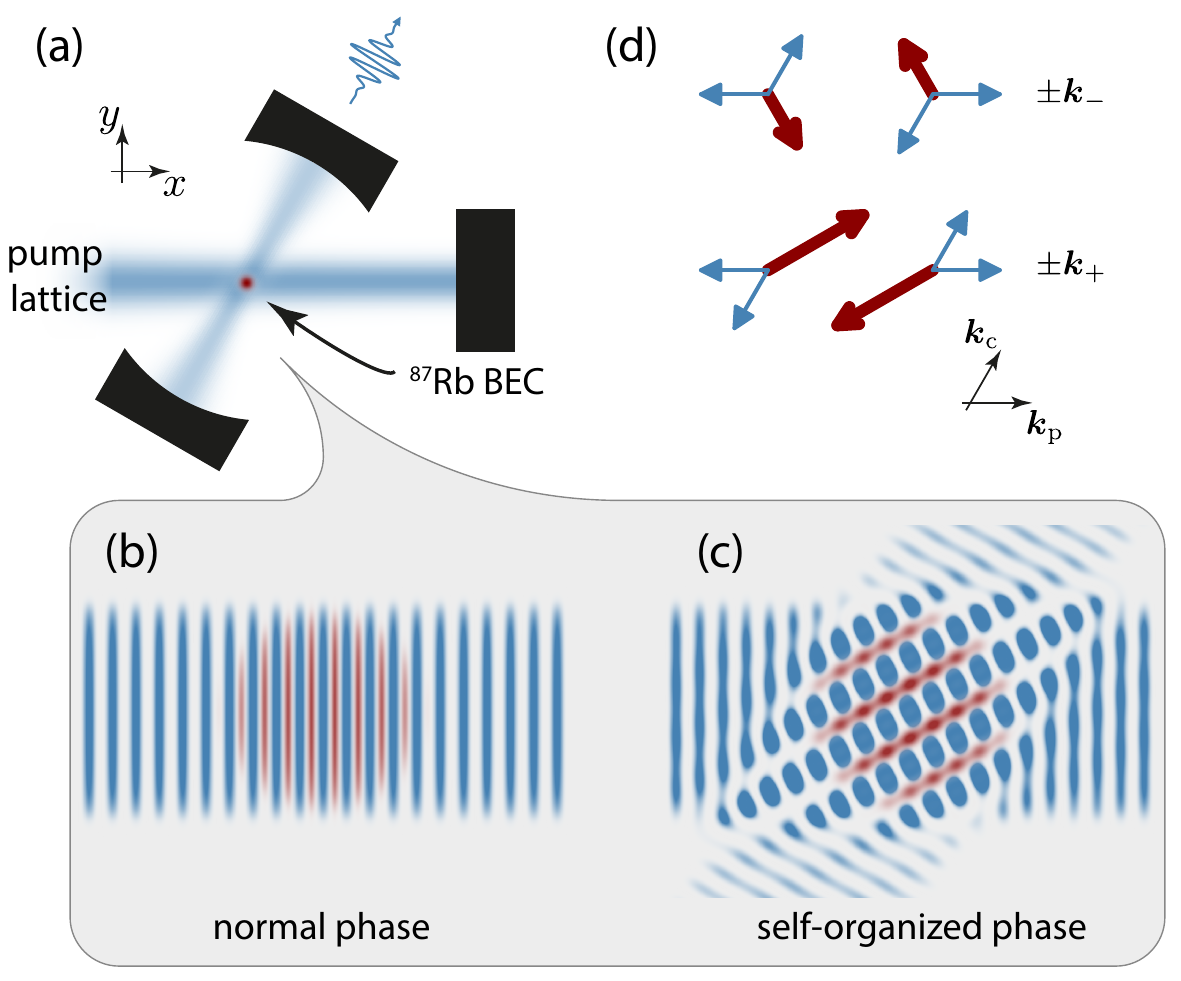}
\caption{(a), (b) We prepare a BEC of $^{87}$Rb atoms in the initially unpopulated mode of an optical resonator, and expose it to a repulsive optical lattice (blue) localizing the atoms (red) in the field minima. (c) Populating the cavity mode with photons can become energetically favorable for the system because interference between the pump and the cavity fields gives the atoms more space to expand close to the nodes of the total field. The atoms form a Bragg grating that allows scattering into the cavity mode. Photons leaking from the cavity are detected. (d) The atoms recoil and acquire momentum (red arrows) from scattering photons (blue arrows). These momenta are $\pm\boldsymbol k_-=\pm(\boldsymbol k_\mathrm{c}-\boldsymbol k_\mathrm{p})$ (upper processes) and $\pm\boldsymbol k_+=\pm(\boldsymbol k_\mathrm{c}+\boldsymbol k_\mathrm{p})$ (lower processes).}
\label{fig1}
\end{figure}

Our experimental setup is depicted in Fig.~1(a). The experiments start with the creation of a BEC of $N=2.7(1)\times10^5$ $^{87}$Rb atoms by optical evaporation in a crossed dipole trap \cite{SM}. The BEC is placed in the mode of an optical resonator with a decay rate of $\kappa=\unit{2\pi\times147(4)}{kHz}$ and atom-cavity coupling $g_0 = \unit{2\pi\times1.95(10)}{MHz}$. We apply the pump lattice beam to the atoms at an angle of $60(1)^\circ$ with respect to the cavity mode. The lattice depth $V_\mathrm{p}$ is ramped up linearly over time in $\unit{50}{ms}$ from 0 to 17(1)\,$E_\mathrm{rec}$, where $E_\mathrm{rec}=  2\pi \hbar \times3.77$\,kHz and $\hbar$ is the reduced Planck's constant. The pump beam has a wave vector $\boldsymbol k_\mathrm{p}=(2\pi/\lambda) \boldsymbol x$ with wave length $\lambda=\unit{780.1}{nm}$, and corresponding frequency $\omega_\mathrm{p}$. It is blue-detuned by $\Delta_\mathrm{a}=\omega_\mathrm{p} - \omega_\mathrm{a} = \unit{2\pi\times76.6(1)}{GHz}$ with respect to the D2 line of $^{87}$Rb at $\omega_\mathrm{a}$ (see \cite{SM} for the effect of changing $\Delta_\mathrm{a}$), and detuned by  $\Delta_\mathrm{c} = \omega_\mathrm{p} - \omega_\mathrm{c}$ with respect to the resonance $\omega_\mathrm{c}$ of the cavity mode with wave vector $\boldsymbol k_\mathrm{c}$. We detect photons leaking from the cavity with a single-photon counting module and convert the detection rate to an intracavity lattice depth. We use this cavity lattice depth as the order parameter of the phase transition.

\begin{figure}
\centering
\includegraphics[width=\columnwidth]{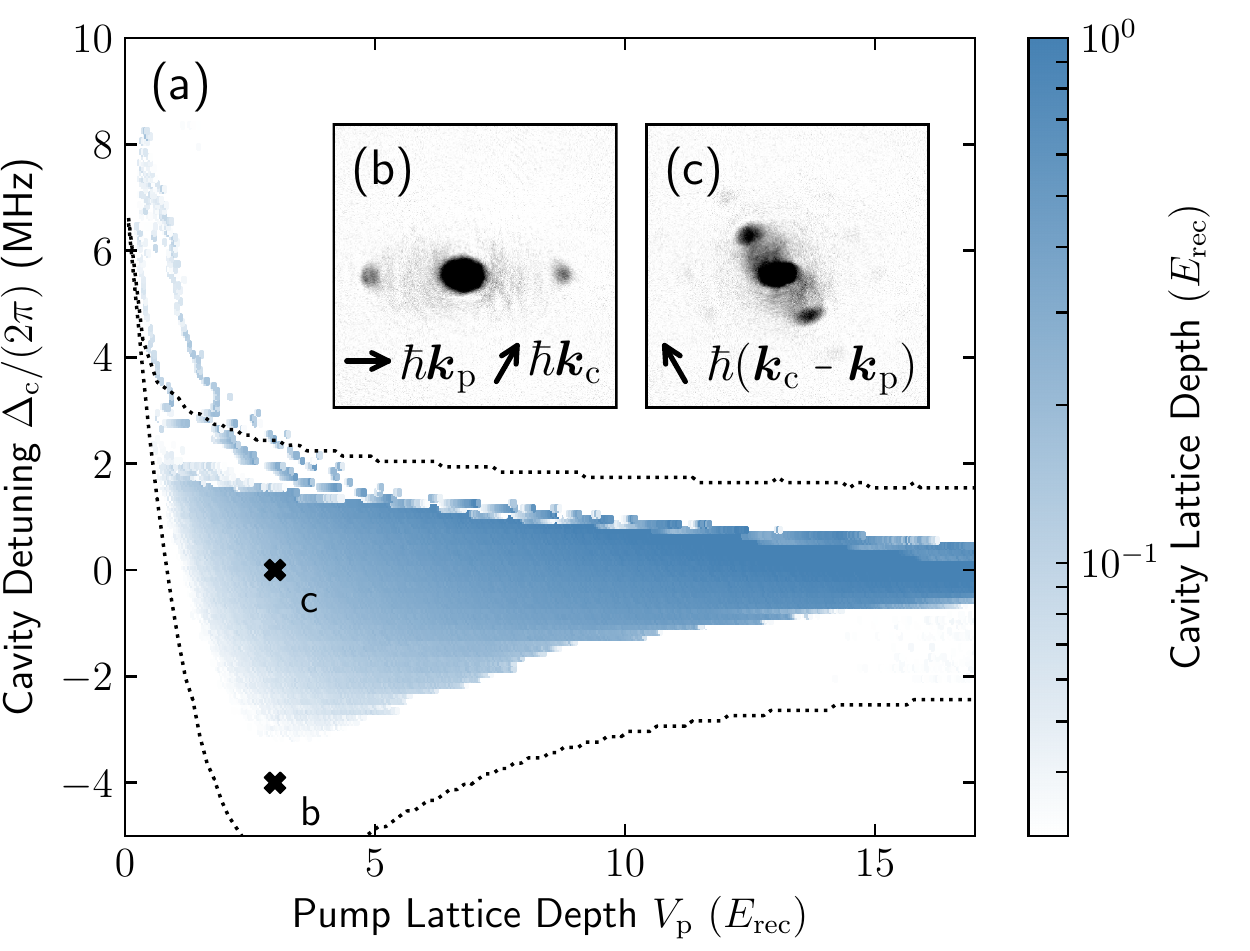}
\caption{Self-organized phase with a repulsive pump. (a) Phase diagram showing the intra-cavity lattice depth as a function of pump lattice depth and cavity detuning. A self-ordered phase with limited extent, identified by a finite lattice in the cavity, can be identified. The dotted lines are the phase boundaries obtained from numerical mean-field calculations. The deviation from the data can be attributed to finite temperatures \cite{SM}. Insets display atomic momentum distributions for a pump lattice of $3.0(2)\,E_\mathrm{rec}$ recorded with absorption imaging after ballistic expansion. (b) For $\Delta_\mathrm{c}=-2\pi\times4.0(1)\,$MHz, the system is in the normal phase with momentum components at 0 and $\pm2\hbar\boldsymbol k_\mathrm{p}$. In real-space, this distribution corresponds to the density modulation of the pump lattice with $\lambda/2$ spacing. (c) At $\Delta_\mathrm{c}=0$, the system is self-organized, and the atoms assume a 1D density modulation with $\lambda$ spacing with momentum components $\pm\hbar(\boldsymbol k_\mathrm{c}-\boldsymbol k_\mathrm{p})$. Both the pump lattice modulation at $\pm2\hbar\boldsymbol k_\mathrm{p}$ and the recoil from the other scattering process at $\pm\hbar(\boldsymbol k_\mathrm{c}+\boldsymbol k_\mathrm{p})$ (see Fig.~1(d)) are suppressed.
}
\label{fig2}
\end{figure}

Repeating this experiment for different cavity detunings $\Delta_\mathrm{c}$, we construct the phase diagram of the system as shown in Fig.~2(a). Blue regions indicate finite mean intra-cavity photon numbers which we identify with a self-ordered phase. For small lattice depths $V_\mathrm{p}$ and  in the vicinity of $\Delta_\mathrm{c}=0$, the phase boundary is approximately linear. Its slope changes for increasing lattice depth, forming a tip of the self-organized phase at a finite cavity detuning $\Delta_\mathrm{c}\approx - \unit{2\pi\times3}{MHz}$, from where the phase boundary bends up again and converges towards $\Delta_\mathrm{c}=0$ for large pump lattice depths. This is in stark contrast to the case of a red-detuned pump lattice ($\Delta_\mathrm{a}<0$), where the phase boundary is monotonic such that a phase transition to the self-ordered phase exists for any $\Delta_\mathrm{c}<0$ \cite{Baumann2010}. For finite positive $\Delta_\mathrm{c}$, we observe a self-ordered phase which however also disappears towards higher pump lattice depths, followed by a few lines in the phase diagram. In the red-detuned case, no self-organization is observed for positive $\Delta_\mathrm{c}$ due to the opposite sign of the dispersive shift.

Absorption images after ballistic expansion of the atoms reveal their momentum distribution. In the normal phase, the atoms are localized in the nodes of the pump lattice (Fig.~2(b)). In the self-ordered phase, the atoms acquire a strong density modulation in the direction $\boldsymbol k_-=\boldsymbol k_\mathrm{c}-\boldsymbol k_\mathrm{p}$ (Fig.~2(c)). This modulation has twice the pump lattice periodicity ($\lambda$) due to the chosen angle of 60$^\circ$ between pump and cavity, the qualitative experimental results however do not depend on this choice. In real-space, this corresponds to a Bragg grating scattering photons from the pump into the cavity. While four different scattering processes are possible as depicted in Fig.~1(d), mostly the scattering processes populating the momenta at $\pm \boldsymbol k_-$ occur (upper graphs in Fig.~1(d)), and the processes populating the momenta at $\pm \boldsymbol k_+ =\pm(\boldsymbol k_\mathrm{c} + \boldsymbol k_\mathrm{p})$ are suppressed. In the self-ordered phase also the density modulation by the pump lattice is suppressed because of the destructive interference reducing the pump potential, as can be seen by the decreased population of the $\pm 2 \hbar \boldsymbol k_\mathrm{p}$ momenta in Fig.~2(c) compared to Fig.~2(b). Again, this is in contrast to the case of a red-detuned pump lattice, where all possible momentum states are macroscopically populated.

\begin{figure*}
\centering
\includegraphics[width=0.8\textwidth]{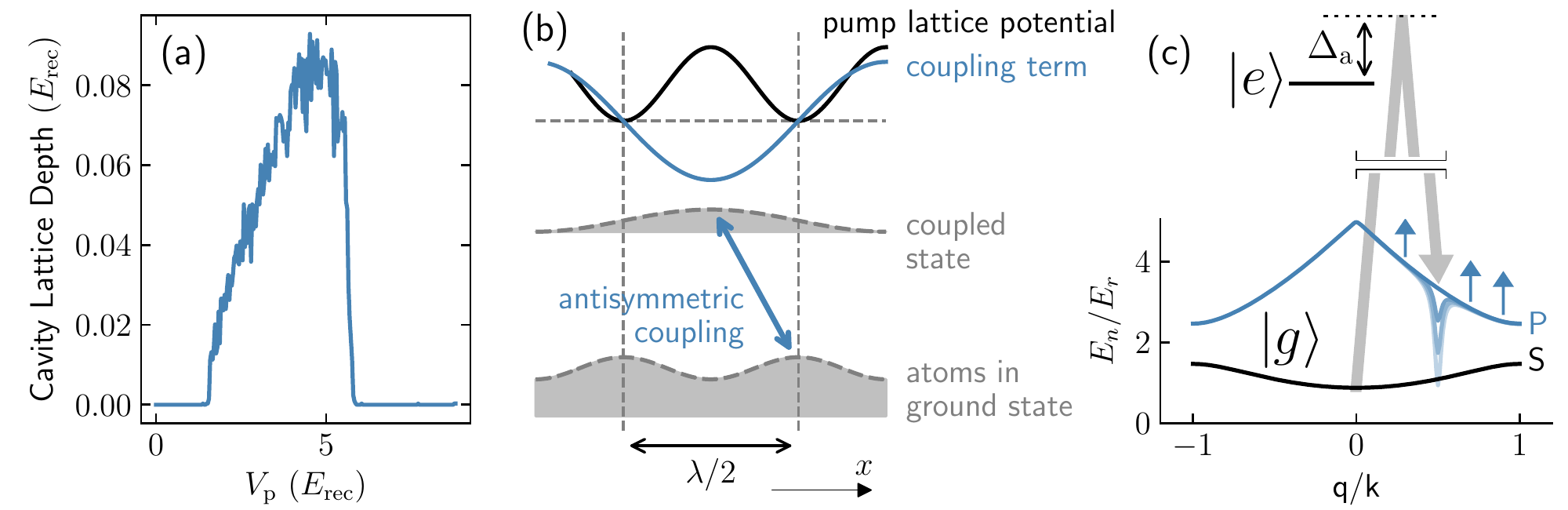}
\caption{Finite extent of the self-organized phase. (a) Cut through the phase diagram for $\Delta_\mathrm{c}=-2\pi\times2$\,MHz showing the cavity lattice depth rising from and returning to 0. (b) The atoms (gray) get localized to the nodes of the repulsive standing wave field (black) along the $x$-direction. Taking the atomic density maxima as origin of the coordinate system, the coupling term with the double period (blue) turns into a sine. It thus couples to a state of opposite symmetry, as indicated. (c) This opposite-symmetry state lives in the P-band of the pump lattice. With increasing pump lattice depth, the coupling term leads to a mode softening at wave vector $\boldsymbol k_-=\boldsymbol k_\mathrm{c}-\boldsymbol k_\mathrm{p}$ (pale blue lines), and finally to self-organization. At the same time, the increasingly deep pump lattice increases the energy of the P-band (blue arrows). For deeper lattices, this effect becomes dominant and leads to the end of the self-ordered phase.}
\label{fig3}
\end{figure*}

These observations can be understood from analyzing the Hamiltonian of the atom-cavity system consisting of the following five terms \cite{Maschler2008}
\begin{equation}
\H = \H_\mathrm{ph} + \H_\mathrm{kin} + \H_\mathrm{pot}^\mathrm{pump} + \H_\mathrm{pot}^\mathrm{cavity} + \H_\mathrm{pot}^\mathrm{coupl}.
\label{eq:H}
\end{equation}
The atoms self-order when the scattering of photons lowers the energy of the system, i.e.~when the gain in potential energy exceeds the energy cost of cavity photons, $\H_\mathrm{ph}=-\hbar \Delta_\mathrm{c} \hat a^\dagger \hat a$, plus the kinetic energy cost caused by recoiling atoms, $\H_\mathrm{kin}=\frac{\hat{p}^2}{2m}$. Here, $\Delta_\mathrm{c}$ is the detuning of the pump frequency relative to the cavity resonance, $\hat a^\dagger$ and $\hat a$ create and annihilate cavity photons, respectively, $\hat p$ is the atomic momentum operator, and $m$ the atomic mass. The potential energy has the three terms
\begin{eqnarray}
\label{eq:pump}
&\H_\mathrm{pot}^\mathrm{pump} &= V_\mathrm{p} \cos^2(\boldsymbol k_\mathrm{p} \cdot  \hat{\boldsymbol r}) \\
\label{eq:cav}
&\H_\mathrm{pot}^\mathrm{cavity} &= U_0 \hat a ^\dagger \hat a \cos^2(\boldsymbol k_\mathrm{c} \cdot \hat{\boldsymbol r}) \\
&\H_\mathrm{pot}^\mathrm{coupl} &= \sqrt{V_\mathrm{p} U_0 }(\hat a + \hat a^\dagger) \cos(\boldsymbol k_\mathrm{c} \cdot \hat {\boldsymbol r}) \cos(\boldsymbol k_\mathrm{p} \cdot \hat {\boldsymbol r}) \,,
\label{eq:coupling}
\end{eqnarray}
where $U_0=0.012\,E_\mathrm{rec}$ is the dispersive cavity shift per atom.     

Eq.~\ref{eq:coupling} describes the coupling of the pump lattice to the light field of the cavity and the atomic density. This leads to a polariton mode, a coupled excitation of the cavity light field and the atomic density at momenta $\pm\boldsymbol k_\pm$. With increasing pump lattice depth, its energy is reduced \cite{Mottl2012}. The polariton mode softens and approaches zero energy at the critical point, manifesting a continuous phase transition.

Although Eq.~\ref{eq:coupling} couples equally to the $\boldsymbol k_-$ and $\boldsymbol k_+$ modes, there is a strong imbalance between their populations in the self-ordered phase (Fig.~2(c)) for two reasons. The first is the difference in their respective kinetic energies of \unit{1}{$E_\mathrm{rec}$} and \unit{3}{$E_\mathrm{rec}$}, respectively. The second, more intricate one is that Eqs.~\ref{eq:pump} and \ref{eq:cav} have finite matrix elements coupling the two modes with $\bra{\boldsymbol k_-}(\H_\mathrm{pot}^\mathrm{pump}+\H_\mathrm{pot}^\mathrm{cavity})\ket{\boldsymbol k_+}\propto1/\Delta_\mathrm{a}$. For $\Delta_\textrm{a}>0$, a population of both modes is increasing the total energy of the system and is thus disfavored, opposite to the case of $\Delta_\textrm{a}<0$.

The finite extent of the self-ordered phase for large pump lattice depths and negative cavity detunings (Fig.~2(a) and Fig.~3(a)) can be explained with a symmetry argument  (see Fig.~3(b)): For deep lattices, the atoms in the ground state localize in the minima of the potential in Eq.~\ref{eq:pump} $\propto \cos^2(\boldsymbol k_\mathrm{p}\cdot\boldsymbol r)$, which for $\Delta_\mathrm{a}>0$ are located at the nodes of the light field. The maxima of the atomic density distribution are thus shifted by $\lambda/4$ with respect to the pump lattice. Accordingly, taking the atomic density maxima as origin of the coordinate system, the coupling term in Eq.~\ref{eq:coupling} becomes $\propto \sin(\boldsymbol k_\mathrm{p}\cdot\boldsymbol r)$. Being an odd operator, it couples the atoms in the ground state to a state of opposite parity. Due to its opposite symmetry, this state is in the P-band of the pump lattice and localized at the maxima of the pump potential. As we ramp up the pump lattice, the coupling that leads to the mode softening  is enhanced (pale blue lines in Fig.~\ref{fig3}(c)), enabling self-organization. But with further increasing pump lattice depth, the band gap increases (blue arrows), which pushes the system out of the self-ordered phase again. The phase boundaries arise from the competition between band gap and mode softening.

The above arguments are confirmed by going to the mean-field limit of Eq.~\ref{eq:H} and solving the steady-state equations in perturbation theory \cite{SM}. We find the criterion for superradiance 
\begin{equation}
1 < \frac{4 N U_0 \tilde{\Delta}_\mathrm{c}}{ \tilde{\Delta}_\mathrm{c}^2+\kappa^2} V_\mathrm{p} \chi_\mathrm{sp} (V_\mathrm{p})\,,
\end{equation}
where we defined the dispersively shifted cavity detuning $\tilde{\Delta}_\mathrm{c}=\Delta_\mathrm{c}-N U_0/2$ and the single-particle response function
\begin{equation}
\chi_\mathrm{sp} (V_\mathrm{p}) = \sum\limits_{n,\boldsymbol k} \frac{|\bra{\psi_n^{(0)}(\boldsymbol k)} \hat \Theta \ket{\psi_1^{(0)}(\boldsymbol0)}|^2}{E_1^{(0)}(\boldsymbol0)-E_n^{(0)}(\boldsymbol k)}.
\label{eq:response}
\end{equation}
$\psi_n^{(0)}(\boldsymbol k)$ are the unperturbed Bloch wave functions in the $n$-th band of the pump lattice with eigenenergies $E_n^{(0)}$, and $\hat \Theta = \cos(\boldsymbol{k}_\mathrm{p} \cdot \hat{\boldsymbol{r}}) \cos(\boldsymbol{k}_\mathrm{c}\cdot \hat{\boldsymbol{r}})$.

Taking into account only the two lowest bands and taking the limit of weak lattices we can evaluate \cite{SM}
\begin{equation}
-V_\mathrm{p} \chi_\mathrm{sp} (V_\mathrm{p}) \approx \frac {V_\mathrm{p}}{3 E_\mathrm{rec}} \left( 1 - \frac{V_\mathrm{p}}{8E_\mathrm{rec}} \right).
\end{equation}
This function first rises to a maximum value at $V_\mathrm{p}=4E_\mathrm{rec}$, which can be enough to enter the self-organized phase depending on $\Delta_\mathrm{c}$, but then decreases leading to the end of the self-ordered phase. This reproduces the form of the lower edge of the ordered phase in Fig.~2(a). This is very different from the familiar self-organization for $\Delta_\mathrm{a}<0$, for which the pump lattice -- coupling the BEC to the lowest band -- reduces the bandwidth $E_1^{(0)}(\boldsymbol k)-E_1^{(0)}(\boldsymbol 0)$ and increases the overlap $|\bra{\psi_1^{(0)}(\boldsymbol k)} \hat \Theta \ket{\psi_1^{(0)}(\boldsymbol0)}|^2$ and thus works in favor of, rather than against, self-organization. 
The repulsive nature of the potentials and the band effects counteracting the mode softening lead to a marginal energy difference between the normal and the self-ordered phase. This energy gain of the ordered phase goes down with detuning so that finite temperatures shift the lower tip in the phase diagram up (cf.~theory lines in Fig.~2(a) and see \cite{SM} for data).

\begin{figure}
\centering
\includegraphics[width=\columnwidth]{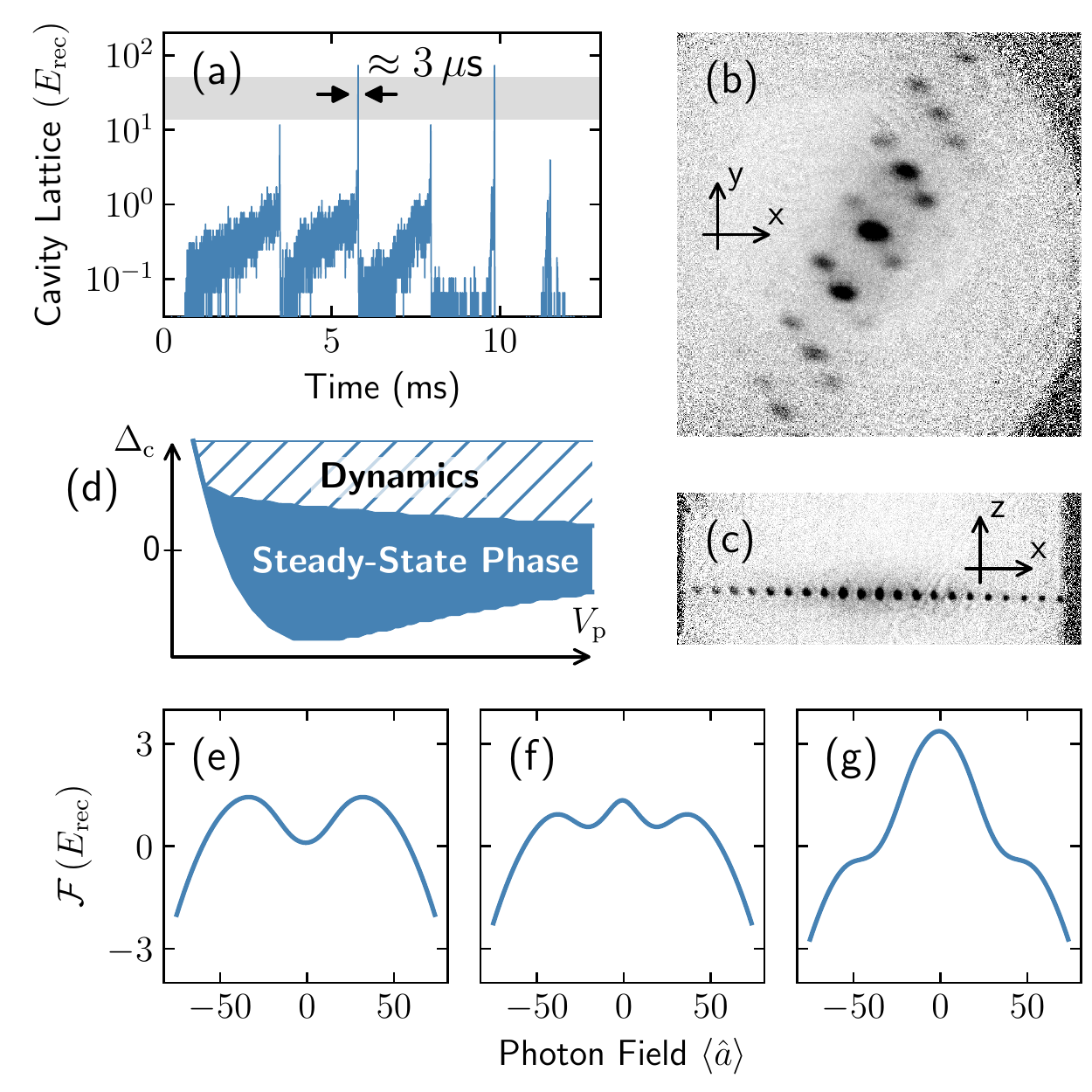}
\caption{Superradiant dynamics. (a) Time-trace of the cavity lattice depth after a far-detuned ramp-up of the pump to 4.1\,$E_\mathrm{rec}$ and a subsequent jump of detuning from $\Delta_\mathrm{c}=2\pi\times -10\,\mathrm{MHz}$ to $\Delta_\mathrm{c}=2\pi\times2\,\mathrm{MHz}$ at time 0. The difference in spike heights stems from the finite photon count rate at which the SPCM saturates \cite{SM}. The gray bar indicates the estimated uncertainty. (b,c) Momentum distribution from absorption imaging after ballistic expansion after a photon spike in the cavity-pump-plane (xy, (b)) and an orthogonal plane (xz, (c)). Very high momenta up to $\pm20\hbar\boldsymbol k_\mathrm{c}$ are visible, possibly limited by the aperture of the imaging system. (d) Schematic phase diagram with stable und dynamic regions. (e-g) Numerical calculations of the free energy per atom as a function of the cavity photon field for $\Delta_\mathrm{c}=2\pi\times$2\,MHz and pump lattice depths of 0.2\,$E_\mathrm{rec}$ (normal phase with local minimum at $\langle\hat a\rangle=0$), 3.4\,$E_\mathrm{rec}$ (self-ordered phase with local minima at a finite photon number) and 13\,$E_\mathrm{rec}$ (spike region, no local minimum).}
\label{fig4}
\end{figure}

Qualitatively different behavior arises for $\Delta_\mathrm{c}>0$ in Fig.~2(a), where we find striking dynamical effects. With increasing pump strength the system enters the self-ordered phase, but then the cavity photon number spikes and the order parameter vanishes. This process re-occurs several times. To disentangle pump from time dynamics, we perform quench experiments in which we ramp up the pump lattice depth at fixed $\Delta_\mathrm{c}=2\pi\times-\unit{10}{MHz}$ in the non-organized phase. We then jump the detuning to the parameter region of interest, and subsequently, keeping pump and detuning constant, record the time evolution of the cavity field. Fig.~4(a) shows an exemplary trace outside the stable self-ordered phase at $\Delta_\mathrm{c}=\unit{2\pi\times2.0(1)}{MHz}$ and $V_\mathrm{p}=4.1(2)\,E_\mathrm{rec}$. We observe repeating pulses with a duration of \unit{3(1)}{\mu s} and a cavity lattice depth of \unit{32(17)}{$E_\mathrm{rec}$}, much deeper than the pump lattice, appearing after an initial buildup time. Taking absorption images of the atomic momentum distribution within a millisecond of a spike reveals the population of very high momentum states (Fig.~4(b,c)), as expected for deep lattices. The timescale of build-up and decay of the cavity field spikes is consistent with our cavity decay rate $\kappa=\unit{2\pi\times147(4)}{kHz}$. We find the same behavior everywhere between the boundary of the steady-state phase and $\Delta_\mathrm{c}=NU_0/2$, leading to the schematic phase diagram of Fig.~4(d).

A basic understanding for the appearance of dynamic features for  $0<\Delta_\mathrm{c}<NU_0/2$ can be gained by neglecting dissipation for the moment and considering the free energy landscape of the system. Fig.~4(e) shows the numerically obtained free energy as a function of the cavity field $\langle \hat a \rangle$ for $\Delta_\mathrm{c}=2\pi \times 2$ MHz and a pre-critical pump power. The term $-\Delta_\mathrm{c} \hat a ^\dagger \hat a$ in Eq.~\ref{eq:H} indicates that the system can lower its energy in the rotating frame with more photons, but the local minimum at $\langle \hat a \rangle = 0$ keeps the system in the normal phase. With increasing pump power, local minima at finite cavity fields develop and the atoms self-organize (Fig.~4(f)). Uniquely for $\Delta_\mathrm{a}>0$ \cite{SM} and in contrast to the case $\Delta_\mathrm{c}<0$, here the self-organized state is metastable. In fact, deeper in the phase the small energy barrier shrinks and is eventually overcome (Fig.~4(g)). Thus the system leaves the metastable state, suddenly increasing the photon number in the cavity. 

Termination and cyclic repetition of the pulses may be qualitatively understood from the dynamic dispersive shift that depends on the overlap between cavity mode and atomic wave function. The strong localization of the atoms at the nodes of the cavity mode during a spike can reduce the dispersive shift until the effective cavity resonance is crossed, terminating self-ordering. The cavity mode empties and the process repeats until heating of the BEC and atom loss finally stop the process. A theoretical description of the observed complex time dynamic phenomena requires a master equation approach, which goes beyond the scope of this work.

We demonstrated the existence of stable and dynamic self-ordered phases of a Bose-Einstein condensate coupled to a high-finesse optical cavity mode via a repulsive optical lattice. A theoretical description of the phase boundaries revealed that an antisymmetric coupling to the P band of the pump lattice induces self-organization. This different lattice geometry for $\Delta_\mathrm{a}>0$ could further lead to a qualitatively different coupling behavior in two- or multi-mode scenarios \cite{Morales2018,Kollar2017}. The observed cycling dynamic appears qualitatively in the phase diagram where limit cycles, chaos, and time crystal behavior were theoretically predicted in related models \cite{Diver2014,Piazza2015, Kessler2019}. We presented a novel approach for experimentally and theoretically exploring time dynamics in driven-dissipative systems \cite{Dogra2019,Chiacchio2019,Carusotto2013,Diehl2010,Buca2019,Kohler2018,Molignini2017}.

\begin{acknowledgments}
We thank Nishant Dogra, Manuele Landini, Francesco Piazza, and Helmut Ritsch for insightful discussions. We acknowledge funding from SNF: project numbers 182650 and 175329 (NAQUAS QuantERA) and NCCR QSIT, from EU Horizon2020: ERCadvanced grant TransQ (project Number 742579) and ITN grant ColOpt (project number 721465), from SBFI (QUIC, contract No. 15.0019), from EPSRC grants EP/P009565/1 and EP/K030094/1, and by an Investigator award of the Simons Foundation.
\end{acknowledgments}

\onecolumngrid
\setcounter{equation}{0}
\setcounter{figure}{0}
\section*{Supplemental Material}

\newcommand{\Er}{E_\text{rec}}
\renewcommand{\a}{\hat{a}}
\newcommand{\ad}{\hat{a}^{\dag }}
\newcommand{\Dc}{\Delta _{\mathrm{c}}}
\newcommand{\Vp}{V_{\mathrm{p}}}
\renewcommand{\H}{\hat{\mathcal{H}}}
\newcommand{\pseudosection}[1]{
\smallskip
\noindent
\textbf{#1}\\\noindent}

\renewcommand{\thefigure}{S\arabic{figure}}
\renewcommand{\figurename}{Supplemental Material Figure}
\renewcommand{\theequation}{S\arabic{equation}}

\section{Experimental details}

\pseudosection{Production of the Bose-Einstein condensate (BEC) inside the optical cavity} 
We perform optical evaporation in an optical dipole trap formed by two orthogonal laser beams at a wavelength of $1064\,\text{nm}$, obtaining an almost pure BEC of $N=2.7(1)\times 10^5$ atoms. The final trapping frequencies are $(\omega_x, \omega_y, \omega_z)=2\pi \times (66(1), 75(1), 133(5))\,\mathrm{Hz}$. 
The BEC is positioned at the center of the fundamental Gaussian mode of an optical cavity. 
We optimize the centering of the atomic cloud by moving the trap position using piezo-eletric mirror mounts and maximizing the dispersive shift of the cavity resonance.  

\pseudosection{Cavity and lasers frequency stabilization} 
The cavity is $2.45\,$mm long and has a decay rate of $\kappa = 2\pi\times 147(4)\,\mathrm{kHz}$.
During all the experiments, the cavity length is actively stabilized using a Pound-Drever-Hall lock-in technique with 830 nm laser light. The power of this beam is low enough such that the residual intracavity lattice depth is $\ll0.1\,\Er$, a value which is negligible with respect to the self-organization interference lattices.
Both the 830 nm lock laser and the 780 nm pump lattice laser are frequency locked on a passively stable transfer cavity, which ensures relative drifts lower than $\kappa$ over one day. 

\pseudosection{Lattice and photon number calibrations} 
We perform Raman-Nath diffraction of the BEC to calibrate both the transverse pump lattice depth and the intracavity lattice depth.
From the intracavity lattice we calculate a detection efficiency of $1.9(2)\%$ for our single photon counting module.

\pseudosection{SPCM Saturation}
Our photon detectors exponentially saturate to a count rate $R_\mathrm{max}=22.3$\,MHz. To obtain the correct photon rate $R_\mathrm{true}$ from the detection rate $R_\mathrm{det}$ we calculate
\begin{equation}
R_\mathrm{true} = \frac{R_\mathrm{max} R_\mathrm{det}}{R_\mathrm{max}-R_\mathrm{det}}.
\label{eq:spcmsat}
\end{equation}
The dynamic range of the SPCM for 1\,$\mu$s integration time is 22. Due to Eq.~\ref{eq:spcmsat}, these 22 levels correspond to highly nonlinearly spaced true count rates. In Fig.~4a of the main text, the difference in height between the first and the second pulse is a single photon.

\begin{figure}[h!]
\centering
\includegraphics[width=0.5\columnwidth]{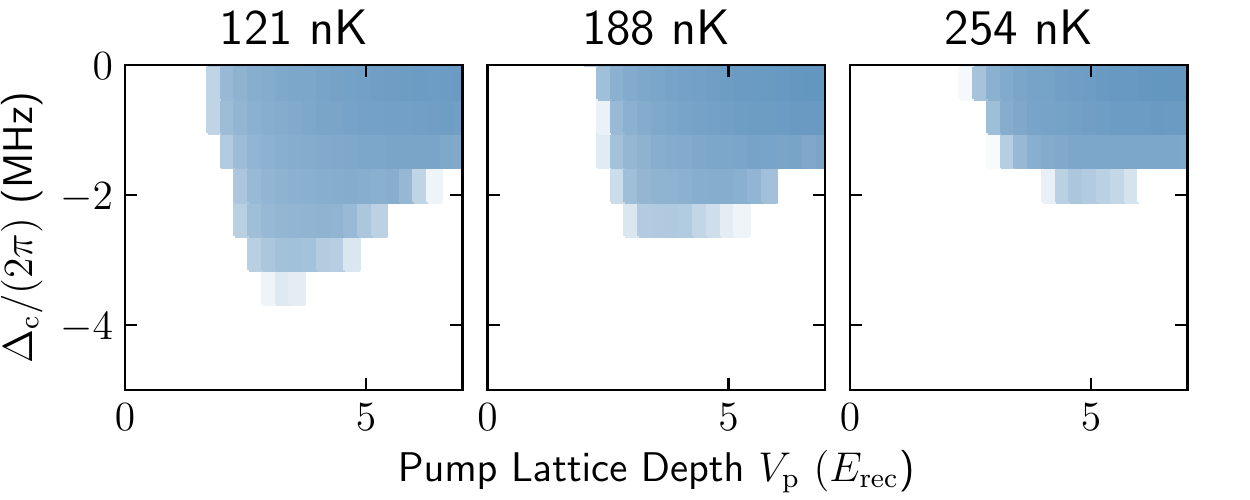}
\caption{These phase diagrams for different atomic temperatures but otherwise identical conditions show that the tip of the self-ordered phase is sensitive to heating. This hints at a low energy gain of the self-ordered phase over the normal phase.}
\label{fig_SM_temperature}
\end{figure}

\pseudosection{Effect of temperature} 
Fig.~\ref{fig_SM_temperature} shows the effect of atomic temperature on the extent of the tip of the self-ordered phase. Higher temperatures push the phase boundaries towards higher pump powers and more positive detunings. 

\begin{figure}[h!]
\centering
\includegraphics[width=\columnwidth]{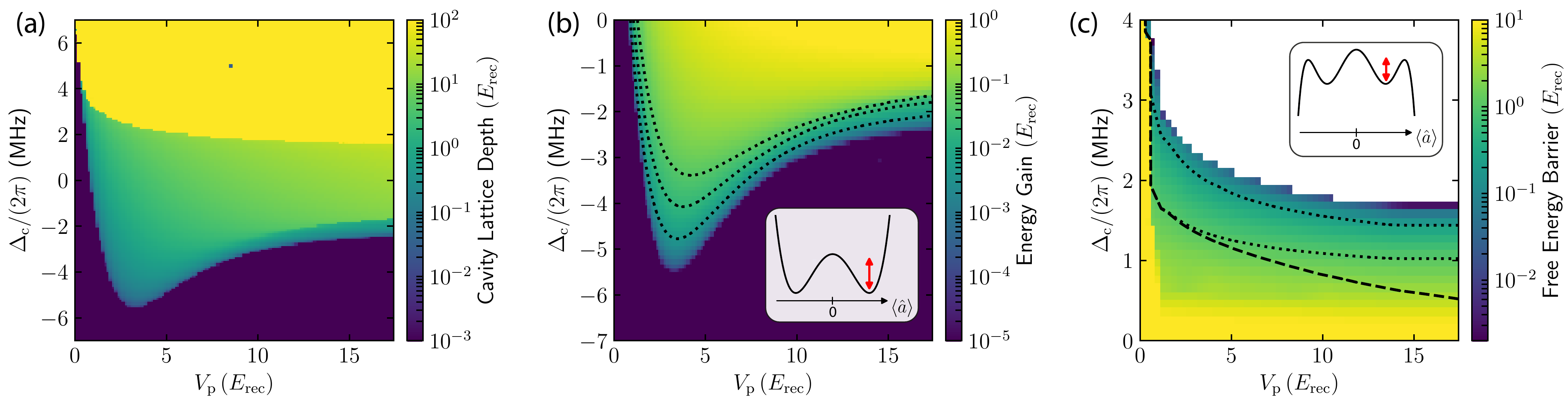}
\caption{(a) At 0 temperature, we define the mean-field phase boundaries by the condition $1<\langle \hat a \rangle<100$ (dotted line in Fig.~2(a) in the main text). (b) To take into account finite temperatures for the lower phase boundary, we calculate the mean-field energy gain of the self-ordered vs.~the normal phase (red arrow in the inset depicting the free energy). Finite temperatures push the phase boundary to higher $\Delta_\mathrm{c}$, as illustrated with the equi-energy contours. (c) For the upper phase boundary, we calculate the barrier of the free energy separating the local minimum from the unbounded region for each point in the phase diagram (red arrow in inset). Finite temperatures bring the phase boundary to lower $\Delta_\mathrm{c}$, again indicated by the dotted equi-energy lines. The dashed line assumes a starting temperature of 180\,nK and a heating rate of 1\,nK/ms/$E_\mathrm{rec}$ pump lattice depth, and agrees with the experimental phase boundary in Fig.~2 within 100\,kHz of $\Delta_\mathrm{c}$. The white region indicates the absence of a finite energy barrier.}
\label{fig_SM_Meanfield_T}
\end{figure}

This matches with mean-field calculations in which we numerically determine the ground state of the Hamiltonian (Fig.~\ref{fig_SM_Meanfield_T}(a)) and its energy. We see that the energy gain of self-organization, defined as the difference in energy between the self-ordered state and one with $\langle\hat a\rangle=0$, is very small in general and particularly in the tip region (Fig.~\ref{fig_SM_Meanfield_T}(b)). For $\Delta_\mathrm{c}>0$ we see that the energy barrier in the free energy as shown in Fig.~4(f) of the main text is small and can be overcome by temperature (Fig.~\ref{fig_SM_Meanfield_T}(c)). 

\begin{figure}[h]
\centering
\includegraphics[width=0.9\columnwidth]{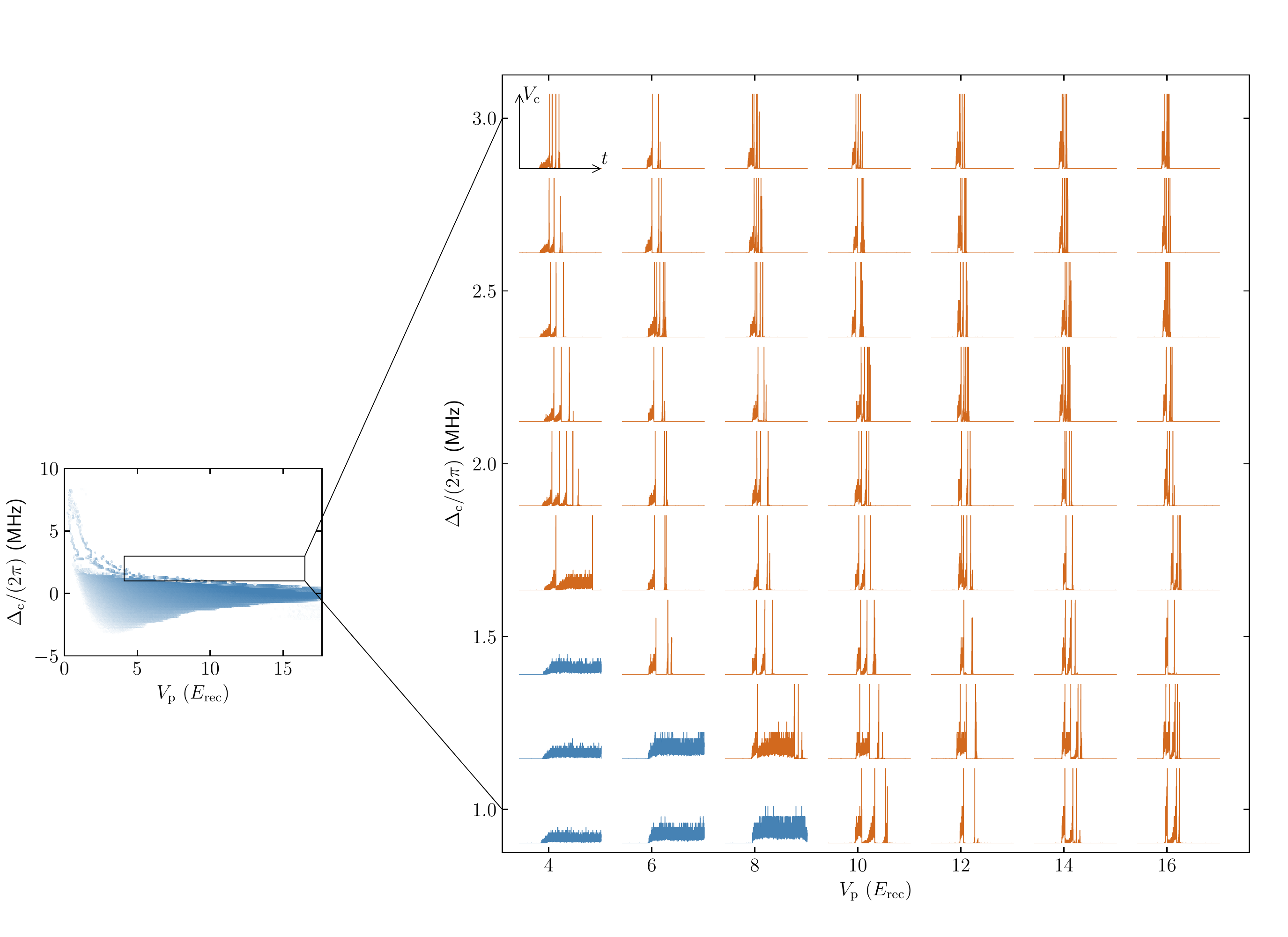}
\caption{Photon traces showing the intra-cavity lattice depth $V_\mathrm{c}$ as a function of time, recorded during 25 ms following our quench protocol. The inset on the phase diagram on the left encompasses the region we explore in this measurement. Each trace on the right corresponds to a different point in the ($V_\mathrm{p}$,$\Dc$) parameter space. All traces share the same linear ordinate axis from 0 to $8\,E_\mathrm{rec}$ (clipping the cycling bursts) \textcolor{black}{and abscissa from 0 to 25\,ms}. In blue and orange color we highlight the traces reaching a steady-state photon level or showing the dynamical behaviour, respectively. The trace $V_\mathrm{p}=4.1\,E_\mathrm{rec}$, $\Delta_\mathrm{c}=2\pi\times2$\,MHz is shown in detail in Fig.~4(a).}
\label{fig_quench}
\end{figure}

\pseudosection{Quench phase diagrams} 
We further explore the dynamical region of the phase diagram in quench experiments. We increase the transverse pump power to different lattice depths $V_\mathrm{p}$ in 5 ms, while the cavity is far detuned ($\Dc \sim -2\pi \times 10$ MHz) in order not to enter the self-organized phase. Then, we suddenly jump the detuning $\Dc$ to the desired value and record the photon level during a holdtime of 25 ms. Repeating the quench for different values of $V_\mathrm{p}$ and $\Dc$ we obtain phase diagrams like the one in Fig.~\ref{fig_quench}. In the steady-state self-organized phase, one can see that after a transient time the cavity field settles to a steady-state value. In contrast, in the dynamical region we record pulses at regular intervals. Furthermore, with respect to Fig. 2(a) of the main text, where we adiabatically ramp the transverse pump power, here one can see that the dynamical phase extends to the whole region of the phase diagram at $V_\mathrm{p}$ values larger than the ones where the first pulse occurs. 
One can also observe that the cycle time of the pulses increases for higher $V_\mathrm{p}$ and $\Dc$ values.

\begin{figure}[h!]
\centering
\includegraphics[width=0.5\columnwidth]{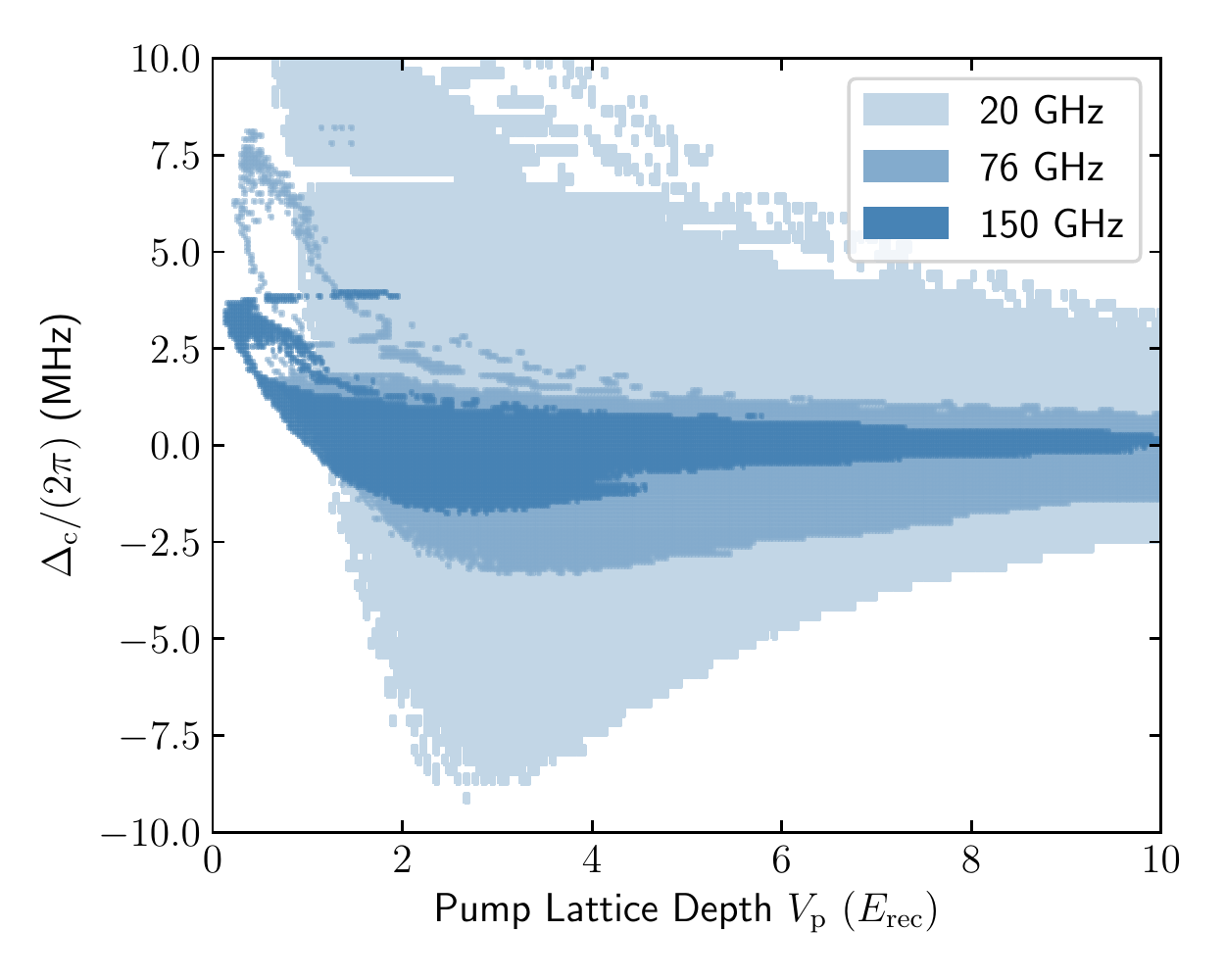}
\caption{Experimentally recorded phase diagrams for different $\Delta_\mathrm{a}$. The vertical extent of the phase diagram scales with $1/\Delta_\mathrm{a}$.}
\label{fig_SM_DeltaA}
\end{figure}

\pseudosection{Different atomic detuning} 
We investigate the effect of the atomic detuning $\Delta_\mathrm{a}$ on the phase diagram presented in Fig.~2 of the main text. The resulting phase diagrams are plotted in Fig.~\ref{fig_SM_DeltaA}.
At fixed $V_\mathrm{p}$, the atomic detuning enters the superradiance condition, Eq.~(5) of the main text, only via the cavity dispersive shift $U_0$. Therefore, we expect a shrinking of the self-organized phase moving further away from atomic resonance. 

\newpage
\section{Theory}

\pseudosection{Criterion for self-organization transition}
The self-organisation phase transition is well understood for red-detuned pump lattices. In the blue-detuned case, the presence of the lattice does not give a simple renormalisation of the critical value, but genuinely modifies the excitation spectrum of the system, giving rise to a finite extent of the self-organized phase.
In the following we will calculate the general result for the critical value of the phase transition and use a perturbative analysis to show that red- and blue-detuned cases indeed are different.
We describe our system in the reference frame of the pump lattice, rotating at a frequency $\omega_\mathrm{p}$. For $N$ particles, we rewrite the Hamiltonian Eq.~(1) from the main text in second quantization formalism as:
\begin{equation}
\begin{split}
\H&=-\hbar\Delta _{\mathrm{c}}\ad\a+\H_{0}\\&+s_{\mathrm{a}}\left[ V_{%
\mathrm{p}}\hat{W}+\ad\a\,U_0 \hat{B}%
+\left( \hat{a}+\hat{a}^{\dag }\right) \sqrt{U_0 V_{\mathrm{p}}}\hat{%
\Theta}\right],
\end{split}
\label{eq:hamiltonianstart}
\end{equation}%
where we made explicit the dependence on the sign of the atomic detuning $\Delta_\mathrm{a}$ via the number $s_{\mathrm{a}} = \text{sign}(\Delta_\mathrm{a})$, to highlight the differences between the red and blue atomic detuning cases. For the experimental results presented in this paper $s_{\mathrm{a}} = + 1$. In Eq.~\eqref{eq:hamiltonianstart} we neglected the effect of the short range $s$-wave inter-particle collisions, $\Dc = \omega_\mathrm{p} - \omega_\mathrm{c}$ is the detuning from the cavity resonance frequency $\omega_\mathrm{c}$, $\ad$ ($\a$) is the creation (annihilation) operator for the cavity photons, $\Vp$ and $U_0$ are the transverse pump and cavity lattice depth respectively and we defined the integrals
\begin{eqnarray*}
\H_{0} &=&\int \hat{\Psi}^{\dag }(\boldsymbol r)\left\{-\frac{1}{2m}\nabla
^{2}\right\} \hat{\Psi}(\boldsymbol r) \,\mathrm{d}\boldsymbol{r} , \\
\hat{W} &=&\int \hat{\Psi}^{\dag }(\boldsymbol r) \cos ^{2}\left( \mathbf{k}_{%
\mathrm{p}}\cdot\mathbf{r}\right) \hat{\Psi}(\boldsymbol r)\,\mathrm{d}\boldsymbol{r}, \\
\hat{B} &=&\int \hat{\Psi}^{\dag }(\boldsymbol r)\cos ^{2}\left( \mathbf{k}_{%
\mathrm{c}} \cdot\mathbf{r}\right) \hat{\Psi}(\boldsymbol r)\,\mathrm{d}\boldsymbol{r}, \\
\hat{\Theta} &=&\int \hat{\Psi}^{\dag }(\boldsymbol r) \cos \left( \mathbf{k}_{%
\mathrm{c}} \cdot\mathbf{r}\right) \cos \left( \mathbf{k}_{\mathrm{p}}\cdot
\mathbf{r}\right)  \hat{\Psi}(\boldsymbol r)\,\mathrm{d}\boldsymbol{r},
\end{eqnarray*}%
where $\hat{\Psi}(\mathbf{r})$ ($\hat{\Psi}(\mathbf{r})^{\dag }$) is the atomic field operator that annihilates (creates) a particle at position $\mathbf{r}$, $m$ is the particle mass, $\mathbf{k}_{\mathrm{p}}$ and $\mathbf{k}_{\mathrm{c}}$ are the transverse pump and cavity momentum, respectively.

Given the different time scales of the atomic and the field dynamics, one can adiabatically eliminate the latter from the Hamiltonian~\eqref{eq:hamiltonianstart} and get the steady state solution for the intra-cavity light $\alpha=\langle \hat a \rangle$ from the Heisenberg equation $0=\mathrm{d}\langle\hat{a}\rangle/\mathrm{d}t=(i/\hbar)\langle[\hat{\mathcal{H}},\hat{a}]\rangle-\kappa\langle\hat{a}\rangle$,
\begin{gather}
\alpha =\frac{Ns_{\mathrm{a}}\sqrt{U_0 V_{\mathrm{p}}}}{\Delta _{%
\mathrm{c}}-s_{\mathrm{a}}U_0 \left\langle \hat{B}\right\rangle  +i\kappa
}\left\langle \hat{\Theta}\right\rangle   ,  \label{SC_EQ} \\
\H_{\mathrm{MF}}\left( \alpha \right) \left\vert \psi \right\rangle
=E\left\vert \psi \right\rangle ,
\end{gather}%
where the mean field single-particle Hamiltonian is given by $\H_{\mathrm{MF}%
}\left( \alpha \right) =\H_{0}+s_{\mathrm{a}}\left[ V_{\mathrm{p}}\hat{W}%
+\left\vert \alpha \right\vert ^{2}U_0 \hat{B}+2\mathrm{Re}\alpha
\sqrt{U_0 V_{\mathrm{p}}}\hat{\Theta}\right] $ \textcolor{black}{and $\vert\psi\rangle$ is the atomic ground state. Note, that -- due to the $\alpha$-dependence of $\vert\psi\rangle$ -- both $\langle \hat{\Theta}\rangle$ and $\langle \hat{B}\rangle$ are functions of $\alpha$}. Equation~\eqref%
{SC_EQ} can be separated into real and imaginary parts as
\begin{eqnarray}
\mathrm{Re}\alpha  &=&\frac{\Delta _{\mathrm{c}}-s_{\mathrm{a}%
}U_{0}\left\langle \hat{B}\right\rangle  }{\left[
\Delta _{\mathrm{c}}-s_{\mathrm{a}}U_{0}\left\langle \hat{B}
 \right\rangle  \right] ^{2}+\kappa ^{2}}Ns_{\mathrm{a}}\sqrt{U_{0}V_{%
\mathrm{p}}}\left\langle \hat{\Theta} \right\rangle  ,
\label{sec_re} \\
\mathrm{Im}\alpha  &=&-\frac{\kappa }{\left[ \Delta _{\mathrm{c}}-s_{\mathrm{a}%
}U_{0}\left\langle \hat{B}\right\rangle  \right]
^{2}+\kappa ^{2}}Ns_{\mathrm{a}}\sqrt{U_{0}V_{\mathrm{p}}}\left\langle \hat{%
\Theta} \right\rangle  .  \label{sec_im}
\end{eqnarray}%
It can be simplified into
\begin{eqnarray}
\mathrm{Re}\alpha  &=&f\left( \alpha \right) ,  \label{sec1_re} \\
\frac{\mathrm{Im}\alpha }{\mathrm{Re}\alpha } &=&-\frac{\kappa }{\Delta _{%
\mathrm{c}}-s_{\mathrm{a}}U_{0}\left\langle \hat{B}
\right\rangle  }.  \label{sec2_im}
\end{eqnarray}%
Here we have defined a function
\begin{equation*}
f\left( \alpha \right) =\frac{\Delta _{\mathrm{c}}-s_{\mathrm{a}%
}U_{0}\left\langle \hat{B}\right\rangle  }{\left[
\Delta _{\mathrm{c}}-s_{\mathrm{a}}U_{0}\left\langle \hat{B}
 \right\rangle  \right] ^{2}+\kappa ^{2}}Ns_{\mathrm{a}}\sqrt{U_{0}V_{%
\mathrm{p}}}\left\langle \hat{\Theta} \right\rangle  .
\end{equation*}%
Note that the condition given by Eq. (\ref{sec2_im}) fixes the phase of the
cavity field. Then the problem is reduced into solving Eq. (\ref{sec1_re}%
) with the condition Eq. (\ref{sec2_im}). Near the critical point, we can
expand Eq. (\ref{sec1_re}) into%
\begin{eqnarray}
\mathrm{Re}\alpha  &=&f\left[ \mathrm{Re}\alpha ,\mathrm{Im}\alpha \left( \mathrm{Re}%
\alpha \right) \right] \\
&=&\left. \frac{df}{d\mathrm{Re}\alpha }\right\vert _{\alpha =0}\mathrm{Re}%
\alpha +\left. \frac{1}{3!}\frac{d^{3}f}{d\left( \mathrm{Re}\alpha \right) ^{3}%
}\right\vert _{\alpha =0}\left( \mathrm{Re}\alpha \right) ^{3}+\cdots .
\end{eqnarray}%
Since $\langle\hat{B}\rangle$ is an even and $\langle\hat{\Theta}\rangle$ an odd function of $\alpha$, $f(\alpha)$ is an odd function and there is no second derivative term. The total derivatives are given by%
\begin{eqnarray*}
\frac{df}{d\mathrm{Re}\alpha } &=&\left( \frac{\partial }{\partial \mathrm{Re}%
\alpha }+\frac{d\mathrm{Im}\alpha }{d\mathrm{Re}\alpha }\frac{\partial }{%
\partial \mathrm{Im}\alpha }\right) f \\
\frac{d^{3}f}{d\left( \mathrm{Re}\alpha \right) ^{3}} &=&\left( \frac{\partial
}{\partial \mathrm{Re}\alpha }+\frac{d\mathrm{Im}\alpha }{d\mathrm{Re}\alpha }%
\frac{\partial }{\partial \mathrm{Im}\alpha }\right) ^{3}f,
\end{eqnarray*}%
where $\frac{d\mathrm{Im}\alpha }{d\mathrm{Re}\alpha }$ is determined by the
condition (\ref{sec2_im})
\begin{equation}
\frac{d\mathrm{Im}\alpha }{d\mathrm{Re}\alpha }=\frac{s_{\mathrm{a}}U_{0}\frac{%
\partial \left\langle \hat{B}\right\rangle  }{\partial
\mathrm{Re}\alpha }\mathrm{Im}\alpha -\kappa }{\Delta _{\mathrm{c}}-s_{\mathrm{a}%
}U_{0}\left\langle \hat{B}\right\rangle  -s_{\mathrm{a}}U_{0}\frac{\partial
\left\langle \hat{B}\right\rangle  }{\partial \mathrm{Im}%
\alpha }\mathrm{Im}\alpha }.
\end{equation}%
To have a non-zero steady solution, one needs%
\begin{eqnarray}
\left. \frac{df}{d\mathrm{Re}\alpha }\right\vert _{\alpha =0} &>&1
\label{cond01} \\
\left. \frac{d^{3}f}{d\left( \mathrm{Re}\alpha \right) ^{3}}\right\vert
_{\alpha =0} &<&0.  \label{cond02}
\end{eqnarray}%
The first condition can be evaluated analytically, giving%
\begin{equation}
\frac{Ns_{\mathrm{a}}\sqrt{U_{0}V_{\mathrm{p}}}\tilde{\Delta}_{\mathrm{c}}}{%
\tilde{\Delta}_{\mathrm{c}}^{2}+\kappa ^{2}} \left.\frac{\partial \left\langle \hat{%
\Theta}\right\rangle  }{\partial \mathrm{Re}\alpha }\right\vert_{\alpha=0}>1
\label{eq:nonzerosolution}
\end{equation}%
where $\tilde{\Delta}_{\mathrm{c}}=\Delta_\mathrm{c}-s_\mathrm{a}U_0\langle B\rangle$ is the dispersively shifted cavity detuning \textcolor{black}{for $\alpha=0$}. The second condition is too complex to be calculated analytically\textcolor{black}{. However,}
it can be numerically computed. In our parameter regime this condition
always holds near the normal phase to superradiant phase transition.

Near the critical point, $%
\mathrm{Re} \alpha $ is very small, such that we can treat the term $2%
\mathrm{Re} \alpha  \sqrt{U_0 V_{\mathrm{p}}}\hat{\Theta}$
as a perturbation. We obtain the perturbed wave functions as
\begin{equation}
\left\vert \psi \right\rangle =\left\vert \psi _{1}^{\left( 0\right) }\left(
\mathbf{0}\right) \right\rangle+2\mathrm{Re} \alpha  \sqrt{%
U_0 V_{\mathrm{p}}}\sum\limits_{m,\mathbf{k}}\frac{ \left\langle
\psi _{m}^{\left( 0\right) }\left( \mathbf{k}\right) \right\vert \hat{\Theta}%
\left\vert \psi _{1}^{\left( 0\right) }\left( \mathbf{0}\right)
\right\rangle }{E_{1}^{\left( 0\right) }\left( \mathbf{0}\right)
-E_{m}^{\left( 0\right) }\left( \mathbf{k}\right) }\left\vert \psi
_{m}^{\left( 0\right) }\left( \mathbf{k}\right) \right\rangle,
\end{equation}%
where the unperturbed wavefunctions $\left\vert \psi
_{m}^{\left( 0\right) }\left( \mathbf{k}\right) \right\rangle$ are the Bloch wave solutions of the unperturbed Hamiltonian $\H_{0}+s_{\mathrm{a}}V_{\mathrm{p}}\hat{W}$, with quasi-momenta $\mathbf{k}$, band index $j$ and eigenenergy $E_{j}^{\left( 0\right) }\left( \mathbf{k}\right)$. The expectation value of $%
\left\langle \hat{\Theta}\right\rangle $ is thus given by%
\begin{equation}
\left\langle \hat{\Theta}\right\rangle =4\mathrm{Re} \alpha
\sqrt{U_0 V_{\mathrm{p}}}\sum\limits_{j,\mathbf{k}}\frac{\left\vert
\left\langle \psi _{j}^{\left( 0\right) }\left( \mathbf{k}\right)
\right\vert \hat{\Theta}\left\vert \psi _{1}^{\left( 0\right) }\left(
\mathbf{0}\right) \right\rangle \right\vert ^{2}}{E_{1}^{\left( 0\right)
}\left( \mathbf{0}\right) -E_{j}^{\left( 0\right) }\left( \mathbf{k}\right) }%
.
\end{equation}%
Defining the response function
\begin{equation}
\chi _{\mathrm{sp}}\left( V_{\mathrm{p}}\right) =\sum\limits_{j,\mathbf{k}}%
\frac{\left\vert \left\langle \psi _{j}^{\left( 0\right) }\left( \mathbf{k}%
\right) \right\vert \hat{\Theta}\left\vert \psi _{1}^{\left( 0\right)
}\left( \mathbf{0}\right) \right\rangle \right\vert ^{2}}{E_{1}^{\left(
0\right) }\left( \mathbf{0}\right) -E_{j}^{\left( 0\right) }\left( \mathbf{k}%
\right) },
\label{eq:responsefunction}
\end{equation}%
one can get from equation~\eqref{eq:nonzerosolution} the superradiance criterion given in the main text as:
\begin{equation*}
\frac{4N U_0 \tilde{\Delta}_{\mathrm{c}}}{\tilde{\Delta}_{\mathrm{c}%
}^{2}+\kappa ^{2}}V_{\mathrm{p}}\chi _{\mathrm{sp}}\left( V_{\mathrm{p}%
}\right) >1.
\end{equation*}

\pseudosection{Weak lattice limit}
The operator $\hat{\Theta}$ in momentum space is given by%
\begin{equation}
\hat{\Theta}=\frac{1}{4}\sum_{\mathbf{k}}\left( \hat{b}_{\mathbf{k}+\mathbf{k%
}_{+}}^{\dag }\hat{b}_{\mathbf{k}}+\hat{b}_{\mathbf{k}-\mathbf{k}_{+}}^{\dag
}\hat{b}_{\mathbf{k}}+\hat{b}_{\mathbf{k}+\mathbf{k}_{-}}^{\dag }\hat{b}_{%
\mathbf{k}}+\hat{b}_{\mathbf{k}-\mathbf{k}_{-}}^{\dag }\hat{b}_{\mathbf{k}%
}\right) .
\end{equation}%
Here $\mathbf{k}_{\pm }=\mathbf{k}_{\mathrm{c}}\pm \mathbf{k}_{\mathrm{p}}$.
The operator $\hat{\Theta}$ can only scatter particles
into four different Bloch wave states $\left( \mathbf{k}_{+},-\mathbf{k}_{+},%
\mathbf{k}_{-},-\mathbf{k}_{-}\right) $. Therefore, the response function can be
simplified into%
\begin{equation}
\chi _{\mathrm{sp}}\left( V_{\mathrm{p}}\right) =4\sum\limits_{j}\frac{%
\left\vert \left\langle \psi _{j}^{\left( 0\right) }\left( \mathbf{k}%
_{-}\right) \right\vert \hat{\Theta}\left\vert \psi _{1}^{\left( 0\right)
}\left( \mathbf{0}\right) \right\rangle \right\vert ^{2}}{E_{1}^{\left(
0\right) }\left( \mathbf{0}\right) -E_{j}^{\left( 0\right) }\left( \mathbf{k}%
_{-}\right) },  \label{response_fun_simple}
\end{equation}%
where we have made use of the symmetry of the band structure,%
\begin{eqnarray*}
\left\vert \psi _{j}^{\left( 0\right) }\left( \mathbf{k}_{+}\right)
\right\rangle &=&\left\vert \psi _{j}^{\left( 0\right) }\left( \mathbf{k}%
_{-}\right) \right\rangle , \\
E_{j}^{\left( 0\right) }\left( -\mathbf{k}_{\pm }\right) &=&E_{j}^{\left(
0\right) }\left( \mathbf{k}_{\pm }\right) , \\
\left\vert \left\langle \psi _{j}^{\left( 0\right) }\left( -\mathbf{k}%
_{-}\right) \right\vert \hat{\Theta}\left\vert \psi _{1}^{\left( 0\right)
}\left( \mathbf{0}\right) \right\rangle \right\vert ^{2} &=&\left\vert
\left\langle \psi _{j}^{\left( 0\right) }\left( \mathbf{k}_{-}\right)
\right\vert \hat{\Theta}\left\vert \psi _{1}^{\left( 0\right) }\left(
\mathbf{0}\right) \right\rangle \right\vert ^{2}.
\end{eqnarray*}

In the limiting case of a weak lattice, $V_{\mathrm{p}}\ll  E_{\mathrm{rec}%
}=\hbar ^{2}\mathbf{k}_{\mathrm{p}}^{2}/(2m)$, we can use perturbation theory to calculate the band structure and show the difference between attractive and repulsive lattices on the self-organisation transition. 
In addition, we truncate the summation in Eq.~\eqref{response_fun_simple}, keeping
only the two lowest bands $j=1,2$. 
First we calculate the eigenenergy and wave function of the ground state%
\begin{eqnarray}
\frac{E_{1}^{\left( 0\right) }\left( \mathbf{0}\right) }{ E_{\mathrm{rec}}}
&\approx &-\frac{1}{2}\left( \frac{V_{\mathrm{p}}}{4 E_{\mathrm{rec}}}\right)
^{2}, \\
\left\vert \psi _{1}^{\left( 0\right) }\left( \mathbf{0}\right)
\right\rangle &\approx &A\left\{ \left\vert \mathbf{0}\right\rangle -\frac{%
s_{\mathrm{a}}V_{\mathrm{p}}}{16 E_{\mathrm{rec}}}\left( \left\vert 2\mathbf{k}%
_{\mathrm{p}}\right\rangle +\left\vert -2\mathbf{k}_{\mathrm{p}%
}\right\rangle \right) \right\}.
\end{eqnarray}%
Second, the energies and the wave functions at $\mathbf{k}_{-}$ are given by%
\begin{eqnarray}
\frac{E_{1}^{\left( 0\right) }\left( \mathbf{k}_{-}\right) }{ E_{\mathrm{rec}}}
&\approx &2-2\cos \theta -\frac{\left( \frac{V_{\mathrm{p}}}{4 E_{\mathrm{rec}}}%
\right) ^{2}}{\cos \theta }, \\
\frac{E_{2}^{\left( 0\right) }\left( \mathbf{k}_{-}\right) }{ E_{\mathrm{rec}}}
&\approx &2+2\cos \theta +\frac{\left( \frac{V_{\mathrm{p}}}{4 E_{\mathrm{rec}}}%
\right) ^{2}}{\cos \theta }, \\
\left\vert \psi _{1}^{\left( 0\right) }\left( \mathbf{k}_{-}\right)
\right\rangle &\approx &A_{1}\left( \left\vert \mathbf{k}_{\mathrm{-}%
}\right\rangle -\frac{\left( \frac{s_{\mathrm{a}}V_{\mathrm{p}}}{4E_{\mathrm{%
r}}}\right) }{4\cos \theta }\left\vert \mathbf{k}_{\mathrm{+}}\right\rangle
\right) , \\
\left\vert \psi _{2}^{\left( 0\right) }\left( \mathbf{k}_{-}\right)
\right\rangle &\approx &A_{1}\left( \left\vert \mathbf{k}_{\mathrm{+}%
}\right\rangle +\frac{\left( \frac{s_{\mathrm{a}}V_{\mathrm{p}}}{4E_{\mathrm{%
r}}}\right) }{4\cos \theta }\left\vert \mathbf{k}_{\mathrm{-}}\right\rangle
\right) ,
\end{eqnarray}%
where $\theta =\pi /3$ is the angle between $\mathbf{k}_{\mathrm{c}}$ and $%
\mathbf{k}_{\mathrm{p}}$, and $\left\vert \mathbf{k}_{\mathrm{c}}\right\vert
=\left\vert \mathbf{k}_{\mathrm{p}}\right\vert $. The coefficients $A_{0}$ and $A_{1}$ are the corresponding normalization constants
\begin{eqnarray*}
A_{0} &=&\frac{1}{\sqrt{1+\frac{1}{8}\left( \frac{V_{\mathrm{p}}}{4E_{%
\mathrm{r}}}\right) ^{2}}} \\
A_{1} &=&\frac{1}{\sqrt{1+\left( \frac{V_{\mathrm{p}}}{4 E_{\mathrm{rec}}}%
\right) ^{2}}}.
\end{eqnarray*}%
Now we can directly obtain the matrix elements as
\begin{eqnarray}
\left\langle \psi _{1}^{\left( 0\right) }\left( \mathbf{k}_{-}\right)
\right\vert \hat{\Theta}\left\vert \psi _{1}^{\left( 0\right) }\left(
\mathbf{0}\right) \right\rangle &=&\frac{1}{4}A_{0}A_{1}\left[ 1-\frac{1}{2}%
\left( \frac{s_{\mathrm{a}}V_{\mathrm{p}}}{4 E_{\mathrm{rec}}}\right) \right] ,
\\
\left\langle \psi _{2}^{\left( 0\right) }\left( \mathbf{k}_{-}\right)
\right\vert \hat{\Theta}\left\vert \psi _{1}^{\left( 0\right) }\left(
\mathbf{0}\right) \right\rangle &=&\frac{1}{4}A_{0}A_{1}\left[ 1+\frac{1}{2}%
\left( \frac{s_{\mathrm{a}}V_{\mathrm{p}}}{4 E_{\mathrm{rec}}}\right) \right] .
\end{eqnarray}%
Then the response function can be calculated as%
\begin{equation}
\chi _{\mathrm{sp}}\left( V_{\mathrm{p}}\right) 
\approx -\frac{1}{3 E_{\mathrm{rec}}}\left[ 1-\frac{1}{2}\left( \frac{s_{%
\mathrm{a}}V_{\mathrm{p}}}{4 E_{\mathrm{rec}}}\right) \right].
\end{equation}%

We obtain%
\begin{eqnarray}
-V_{\mathrm{p}}\chi _{\mathrm{sp}}\left( V_{\mathrm{p}}\right) &\approx&\frac{V_{%
\mathrm{p}}}{3 E_{\mathrm{rec}}}\left( 1+\frac{V_{\mathrm{p}}}{8 E_{\mathrm{rec}}}%
\right) ,\text{ for } \Delta_\mathrm{a}<0 \\
-V_{\mathrm{p}}\chi _{\mathrm{sp}}\left( V_{\mathrm{p}}\right) &\approx&\frac{V_{%
\mathrm{p}}}{3 E_{\mathrm{rec}}}\left( 1-\frac{V_{\mathrm{p}}}{8 E_{\mathrm{rec}}}%
\right) ,\text{ for } \Delta_\mathrm{a}>0
\end{eqnarray}%
Since the response function of the ground state is negative, self-organization will occur in the effective red cavity detuning regime, $%
\tilde{\Delta}_{\mathrm{c}}<0$. In the atomic red detuning case, $-V_{\mathrm{p}%
}\chi _{\mathrm{sp}}\left( V_{\mathrm{p}}\right) $ is a monotonically
increasing function. So when increasing the pumping strength, $-V_{\mathrm{p}%
}\chi _{\mathrm{sp}}\left( V_{\mathrm{p}}\right) $ will exceed the critical
value and trigger the superradiant phase transition for any $\tilde\Delta_\mathrm{c}<0$. However, in the atomic blue detuning regime, $-V_{\mathrm{p}}\chi _{\mathrm{sp}}\left( V_{\mathrm{p}%
}\right) $ will first raise from zero, reach a maximum, and then decrease again. In this situation, the system will first enter into the superradiant phase, and go back to the normal phase at large pump lattice depths.

\pseudosection{Local minimum for $\Dc>0$}
We will now consider an additional feature of the blue detuned pump lattices. In the case of $\Dc>0$, we observe that self-organisation is possible in a metastable regime, where a local minimum appears in the energy of the system. However, when moving to higher pump powers the system suddenly jumps out of the metastable state, resulting in a burst of cavity light.
To study this local minimum, we can investigate the case with $V_{\mathrm{p}}\approx0$. In this
situation, the Hamiltonian is reduced to%
\begin{equation}
\H=-\hbar \Delta _{\mathrm{c}}\hat{a}^{\dag }\hat{a}+\H_{0}+s_{\mathrm{a}}U_0
\hat{B}\left( \hat{a}^{\dag }\hat{a}\right) ,
\end{equation}%
Then the energy as a function of the intra-cavity light field $\alpha $ reads:%
\begin{equation}
E\left( \alpha \right) =-\textcolor{black}{\hbar}\Delta _{\mathrm{c}}\left\vert \alpha \right\vert
^{2}+NE_{1}^{\left( 0\right) }\left( \mathbf{0}\right) ,
\end{equation}%
where $E_{1}^{\left( 0\right) }\left( \mathbf{0}\right) $ is the single-particle ground state energy of the mean field band structure%
\begin{equation}
\left[ \H_{0}+s_{\mathrm{a}}U_0 \left\vert \alpha \right\vert ^{2}\hat{B}%
\right] \left\vert \psi _{1}^{\left( 0\right) }\left( \mathbf{0}\right)
\right\rangle =E_{1}^{\left( 0\right) }\left( \mathbf{0}\right) \left\vert
\psi _{1}^{\left( 0\right) }\left( \mathbf{0}\right) \right\rangle .
\end{equation}%
Depending on the strength of the intra-cavity light field, \textcolor{black}{two different regimes can be reached}.  
Considering the small $\alpha $ limit, $U_0  \left\vert \alpha
\right\vert ^{2}\ll  E_{\mathrm{rec}}$, $E_{1}^{\left( 0\right) }\left( \mathbf{%
0}\right) $ can be calculated perturbatively as%
\begin{equation}
E_{1}^{\left( 0\right) }\left( \mathbf{0}\right) \approx \frac{1}{2}s_{%
\mathrm{a}}U_0  \left\vert \alpha \right\vert ^{2}-2\frac{\left( \frac{1}{%
4}U_0 \left\vert \alpha \right\vert ^{2}\right) ^{2}}{4 E_{\mathrm{rec}}}.
\end{equation}%
While in the large $\alpha $ limit, $U_0  \left\vert \alpha \right\vert
^{2}\gg  E_{\mathrm{rec}}$, $E_{1}^{\left( 0\right) }\left( \mathbf{0}\right) $
can be approximate with a linear function of $U_0 \left\vert \alpha \right\vert ^{2}$%
:%
\begin{equation}
E_{1}^{\left( 0\right) }\left( \mathbf{0}\right) \approx \frac{1}{2}s_{%
\mathrm{a}}U_0 \left\vert \alpha \right\vert ^{2}-c U_0  \left\vert
\alpha \right\vert ^{2},
\end{equation}%
where $c$ is a positive constant. The total energy is then given by:%
\begin{equation}
E\left( \alpha \right)  \approx  
\begin{cases} 
\left( -\Delta _{\mathrm{c}}+\frac{1}{2}%
Ns_{\mathrm{a}}U_0 \right) \left\vert \alpha \right\vert ^{2}-\frac{%
NU_0  ^{2}\left\vert \alpha \right\vert ^{4}}{32 E_{\mathrm{rec}}}, & \left\vert \alpha \right\vert ^{2}\ll \frac{ E_{\mathrm{rec}}}{U_0} , \\
\left( -\Delta _{\mathrm{c}}+\frac{1}{2}%
Ns_{\mathrm{a}}U_0 -NcU_0  \right) \left\vert \alpha \right\vert ^{2} , & \left\vert \alpha \right\vert ^{2}\gg \frac{ E_{\mathrm{rec}}}{\lambda}.
 \end{cases}
\end{equation}%
We note that in the case of
\begin{equation}
0<-\Delta _{\mathrm{c}}+\frac{1}{2}Ns_{\mathrm{a}}U_0 <NcU_0 
\end{equation}%
the energy functional has no lower bound, but has a local minimum at $%
\left\vert \alpha \right\vert =0$. This condition can only appear in the
blue atomic detuning case. 
Contrary to the blue detuned case, for the red atomic detuning there will be no steady state at $\Delta _{\mathrm{c}}>0$. 

In the dispersive shift picture, the effective cavity detuning is given by:%
\begin{equation}
\tilde{\Delta}_{\mathrm{c}}=\Delta _{\mathrm{c}}-\left\langle \psi
_{1}^{\left( 0\right) }\left( \mathbf{0}\right) \right\vert \left( s_{%
\mathrm{a}}U_0 \hat{B}\right) \left\vert \psi _{1}^{\left( 0\right)
}\left( \mathbf{0}\right) \right\rangle
\end{equation}%
With%
\begin{equation*}
\left\langle \psi _{1}^{\left( 0\right) }\left( \mathbf{0}\right)
\right\vert \left( s_{\mathrm{a}}U_0 \hat{B}\right) \left\vert \psi
_{1}^{\left( 0\right) }\left( \mathbf{0}\right) \right\rangle =\frac{%
\partial E_{1}^{\left( 0\right) }\left( \mathbf{0}\right) }{\partial
\left\vert \alpha \right\vert ^{2}},
\end{equation*}%
one can obtain%
\begin{equation}
\left\langle \psi _{1}^{\left( 0\right) }\left( \mathbf{0}\right)
\right\vert \left( s_{\mathrm{a}}U_0  \hat{B}\right) \left\vert \psi
_{1}^{\left( 0\right) }\left( \mathbf{0}\right) \right\rangle  \approx 
\begin{cases}
\frac{1}{2}Ns_{\mathrm{a}}U_0  -\frac{NU_0  ^{2}\left\vert \alpha
\right\vert ^{2}}{16 E_{\mathrm{rec}}},       & \left\vert \alpha
\right\vert ^{2}\ll  E_{\mathrm{rec}}/U_0  \\
\frac{1}{2}Ns_{\mathrm{a}}U_0  -NcU_0 , & \left\vert \alpha \right\vert ^{2}\gg  E_{\mathrm{rec}}/U_0\,.
\end{cases}
\end{equation}%
Here we note that once $\left\langle \psi _{1}^{\left( 0\right) }\left(
\mathbf{0}\right) \right\vert \left( s_{\mathrm{a}}U_0 \hat{B}\right)
\left\vert \psi _{1}^{\left( 0\right) }\left( \mathbf{0}\right)
\right\rangle >\Delta _{\mathrm{c}}$, the system will enter the unstable
regime.

\end{document}